# The Atwood machine revisited using smartphones

## Martín Monteiro[(a)], Cecilia Stari[(b)], Cecilia Cabeza[(c)], Arturo C. Marti[(d),]

[(a)] Universidad ORT Uruguay; monteiro@ort.edu.uy
[(b)] Universidad de la República, Uruguay, cstari@fing.edu.uy
[(c)] Universidad de la República, Uruguay, cecilia@fisica.edu.uy
[(c)] Universidad de la República, Uruguay, marti@fisica.edu.uy

The Atwood machine is a simple device used for centuries to demonstrate the Newton's second law. It consists of two supports containing different masses joined by a string. Here, we propose an experiment in which a smartphone is fixed to one support. With the aid of the built-in accelerometer of the smartphone the vertical acceleration is registered. By redistributing the masses of the supports, a linear relationship between the mass difference and the vertical acceleration is obtained. In this experiment, the use of a smartphone contributes to enhance a classical demonstration.

**Theory**

The Atwood machine is a simple device invented in 1784 by the English mathematician George Atwood[1-3]. It consists of two objects of mass $m_A$ and $m_B$, connected by an inextensible massless string over an ideal massless pulley[1]. Applying the Newton's second law to each mass we obtain

$$m_A g - T = m_A a \qquad (1)$$

$$T - m_B g = m_B a$$

where $g$ is the gravitational acceleration, $T$ is the tension force, and $a$ is the vertical acceleration. Eliminating the tension between these equations we obtain

$$a = \frac{m_A - m_B}{m_A + m_B} g \qquad (2)$$

or in terms of the mass difference $\Delta m$ and the total mass $M$

$$a = \frac{\Delta m}{M} g \qquad (3)$$

As mentioned in the original Atwood's book, many possible experiments can be implemented using his machine[1]. One of the simplest possibilities, adopted here, is, keeping the total mass, to vary $\Delta m$ by redistributing a set of weights. In this case, a linear relationship between the vertical acceleration and the mass difference is obtained.

**The experiment**

In our experimental setup, shown in Fig. 1, an Atwood machine was built using two pulleys. In this figure, it can be appreciated, on the right, a smartphone on the right support (A) and, on the left the support (B) where up to 5 weights can be placed. The smartphone, a LG G2, is kept fixed to the string s as indicated in Fig. 2 using a clamp similar to those provided with tripods or monopod. In this experience, the smartphone is located in such a way that the only relevant axis is the *x*-axis which coincides with the vertical direction.

Initially, the support A contains only the smartphone (the mass of the smarphone and support is $m_A$ = 191.2(2) g) while the support B holds 5 different weights (the mass of the support and the weights is $m_B$ = 190.8(2) g). The total mass, $M$ = 382.0(4) g, is kept constant along the experiment. The system is released and the App Vernier Graphical Analysis[4], shown in Fig. 3, is used to record the acceleration values during the interval in which the support A is going downward and the support B upwards. Of course, care should be taken to avoid hitting the smartphone against the floor.

Once the smartphone is stopped, the App is paused and a plot of the acceleration as a function of time exhibiting a region of constant acceleration or *plateau* is displayed on the screen (see the left panel of Fig. 4). The vertical acceleration and its error are obtained from the mean value and the standard deviation provided by the APP (Fig. 4). The acceleration measured includes the gravitational contribution, so it is necessary to subtract it to obtain the real acceleration[5,6]. Subsequently, each of the weights are removed from the support B and placed on the A. In this way, the mass of the system remains constant, and only the mass difference $\Delta m = m_A - m_B$ is varied. For each configuration the vertical acceleration is measured. Then, in Fig. 5, plotting the acceleration as a function of the mass difference we obtain a straight line whose slope corresponds to g/M as indicated in Eq. 3.

**Analysis and conclusion**

From the slope fitted in the linear regression we obtain a value for the total mass of the system

$$M_{exp} = \frac{g}{pend}$$, which results $M_{exp}$ = (387 ± 20) g. This value is in considerable agreement with the value obtained by direct weighting $M$ = (382.0 ± 0.4) g. We conclude that, thanks to the aid of

the accelerometer of a smartphone it is possible to foster this demonstration and obtain a precise verification of the Newton's second law.

**References**:

[1] Thomas B. Greenslade Jr., "Atwood's machine," Phys. Teach. 23, 24 (1985)

[2] Alexsandra Siqueira, Aparecida do C.S.Almeida, Jaime Frejlich, "Máquina de Atwood," Rev. Bras. Ens. Fis. 21, 95 (1999).

[3] Gordon O. Johnson, "Making Atwood's machine "work"," Phys. Teach. 39, 154 (2001).

[4] App Vernier Graphical Analysis,
https://play.google.com/store/aps/details=id=com.vernier.andriod.graphical.analysis

[5] Patrik Vogt and Jochen Kuhn "Analyzing free fall with a smartphone acceleration sensor," Phys. Teach. 50, 182 (March 2012).

[6] Martin Monteiro, Cecilia Cabeza, and Arturo C Martí, "Acceleration Measurements Using Smartphone Sensors: Dealing with the Equivalence Principle," Rev. Bras. Ens. Fís, 37, x, (2015).

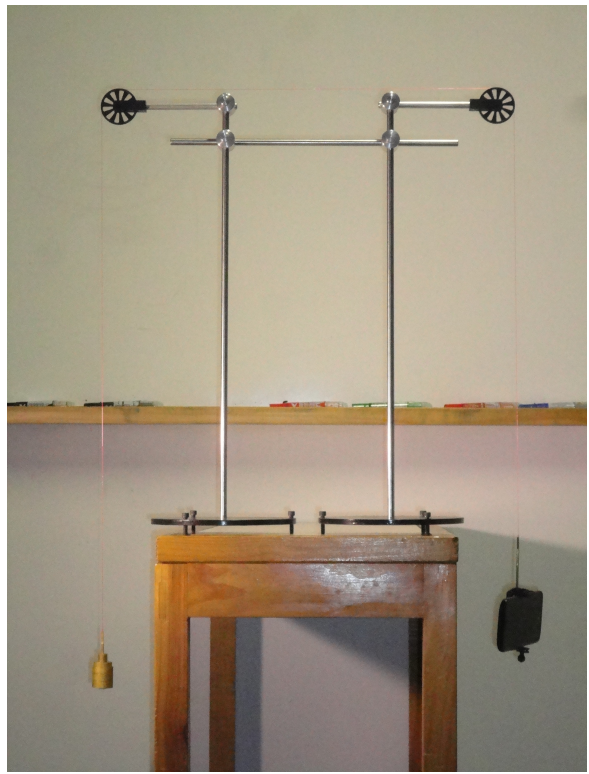

Figure 1. Experimental setup consisting of two supports (A) and (B), connected by a string and supported on two pulleys.

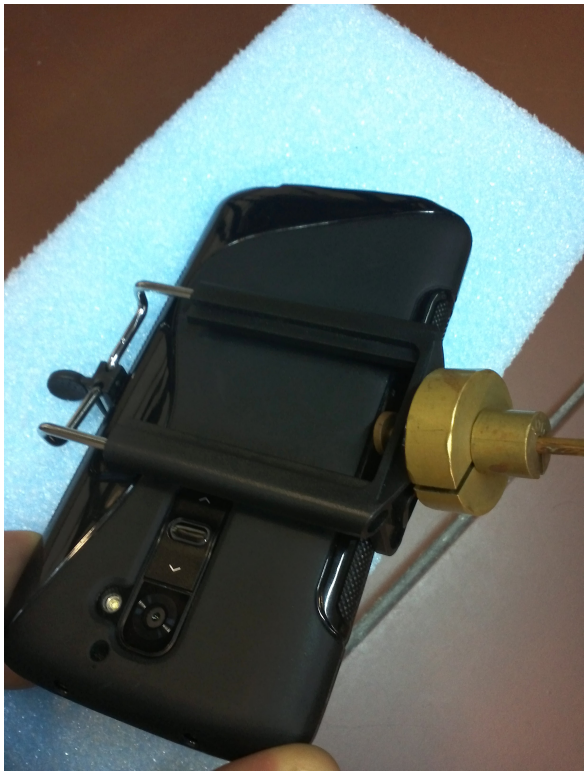

Figure 2. Detail of the support and the clamp.

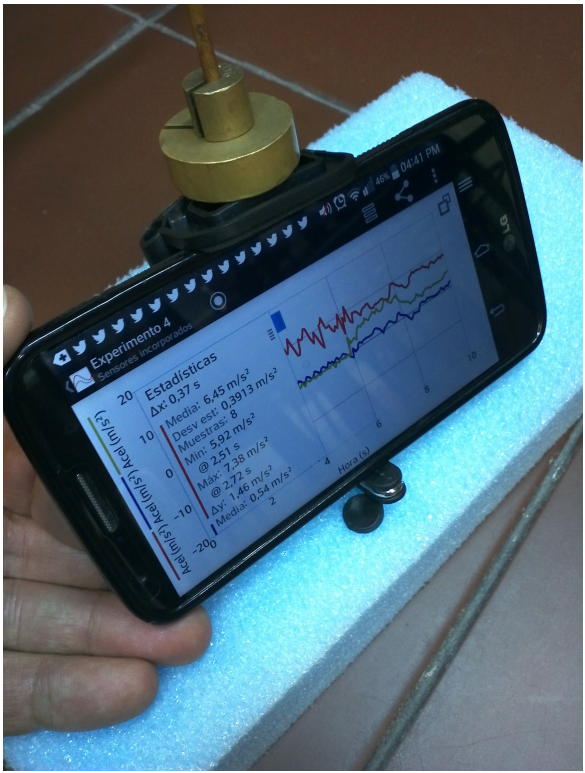

Figure 3. Smartphone mounted on the support showing the App Vernier on the screen.

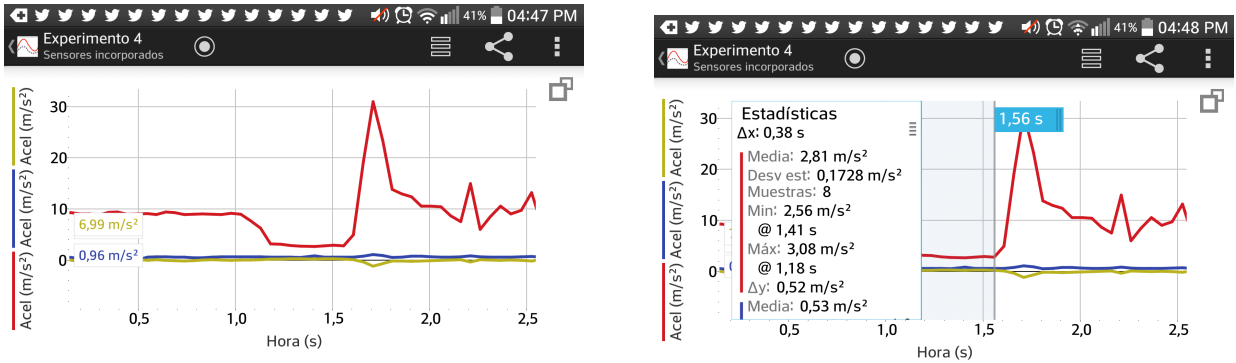

Figure 4. Snapshots of the App Vernier showing the values registered by the acceleration sensor as a function of time. The only relevant component here is the *x* which corresponds to the vertical acceleration. On the left panel, the plateau of constant acceleration can be appreciated. The statistical values calculated in this interval are indicated on the right panel.

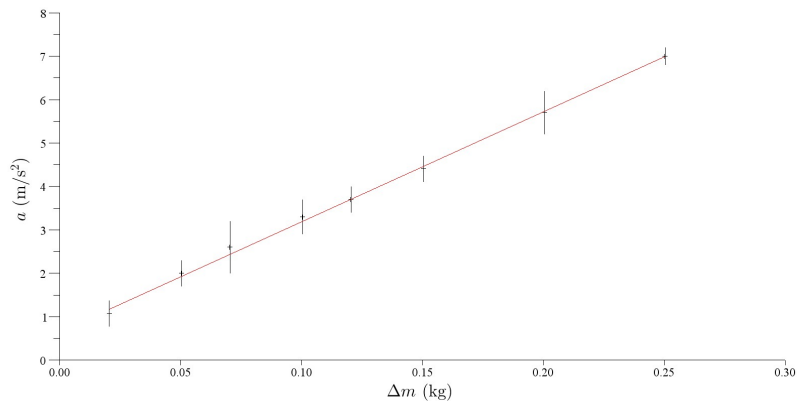

Figure 5. Acceleration as a function of the mass difference: experimental points with error bars (black) and a linear regression (red line).